\documentclass{moriond}

\usepackage[utf8]{inputenc}
\def\be{\begin{equation}}
\def\ee{\end{equation}}
\def\bea{\begin{eqnarray}}
\def\eea{\end{eqnarray}}

\newcommand{\Photo}{}

\begin{document}
\vspace*{4cm}
\title{Cosmic Expansion: A mini review of the Hubble-Lema\^itre tension }

\author{ Francis-Yan Cyr-Racine}

\address{Department of Physics and Astronomy, University of New Mexico, 210 Yale Blvd NE,\\
Albuquerque, NM 87106 USA}

\maketitle

\abstracts{We present here a lightning review of the status of the Hubble-Lema\^itre tension. Instead of discussing the broad array of proposed solutions found in the literature, we focus here on the assumptions made to measure the Hubble constant from cosmic microwave background and baryon acoustic oscillation data on the one hand, and from a calibrated distance ladder on the other hand. From this discussion, we extract two important lessons that inform which kind of physics-based solutions could plausibly resolve this tension.}

\section{Introduction}\label{sec:intro}
By the end of this decade, we will be celebrating the centennial of the discovery of the expansion of the Universe. Ever since the pioneering work of Slipher \cite{1915PA.....23...21S}, Lema\^itre \cite{1927ASSB...47...49L}, Robertson \cite{Robertson:1928}, and Hubble \cite{1929PNAS...15..168H}, the value of the current expansion rate of the Universe has fascinated and puzzled astronomers. Early estimates placed this Hubble-Lema\^itre expansion rate near $H_0\sim 500$ km/s/Mpc, which resulted in a rather awkward universe that is younger than some of the stars it contained \cite{1930MNRAS..90..668E,1933ASSB...53...51L}. Later estimates \cite{1958ApJ...127..513S,1984Natur.307..326S}  pushed the value of the current expansion rate down to between 50 and 100 km/s/Mpc, somewhat alleviating the tension with stellar ages.  

In the current era of precision cosmology, the debate \cite{Verde:2019ivm,Efstathiou:2020wxn,divalentino2021realm} still rages on about what is the exact value of $H_0$, albeit over a smaller range of values. We now have multiple independent ways to infer the value of the expansion rate using an array of cosmological and astrophysical data. Of particular importance for this mini review are the detailed measurements of the temperature and polarization spectra of the cosmic microwave background (CMB) \cite{Hinshaw:2012aka,planck15,Aghanim:2018eyx}, the baryon acoustic oscillation feature imprinted in the large-scale distribution of galaxies \cite{Alam:2016hwk}, and the absolute magnitude of Type Ia supernovae  calibrated using either cepheids \cite{riess11,Riess:2016jrr,Riess:2019cxk} or tip-of-the-red-giant-branch (TRGB) stars \cite{Freedman_2019,Freedman_2020}.

These proceedings briefly summarize the status of the Hubble-Lema\^itre tension as of early 2021. The presentation on which these proceedings are based was aimed at an audience of particle physicists who might not be entirely familiar with the details of how $H_0$ is extracted from both distance-ladder and CMB measurements. The presentation was limited to 15 minutes, which means that I could discuss only a few facets of what is in reality a very rich problem. Thus, instead of covering the wide range of proposed solutions found in the literature, I instead focused on two simple lessons that are relevant to anyone trying to build a successful particle model alleviating the tension. I apologize in advance for not mentioning the vast body of literature that this tension has generated. I also apologize to the hard-working astronomers whose technique to measure $H_0$ I did not have time to mention in the talk: strong gravitational lensing cosmography, cosmic chronometers, gravitational waves, surface brightness fluctuations, Tully-Fisher relation, etc.

\section{The Tension}\label{sec:tension}

\begin{figure}[t]
\begin{center}
\includegraphics[width=\textwidth]{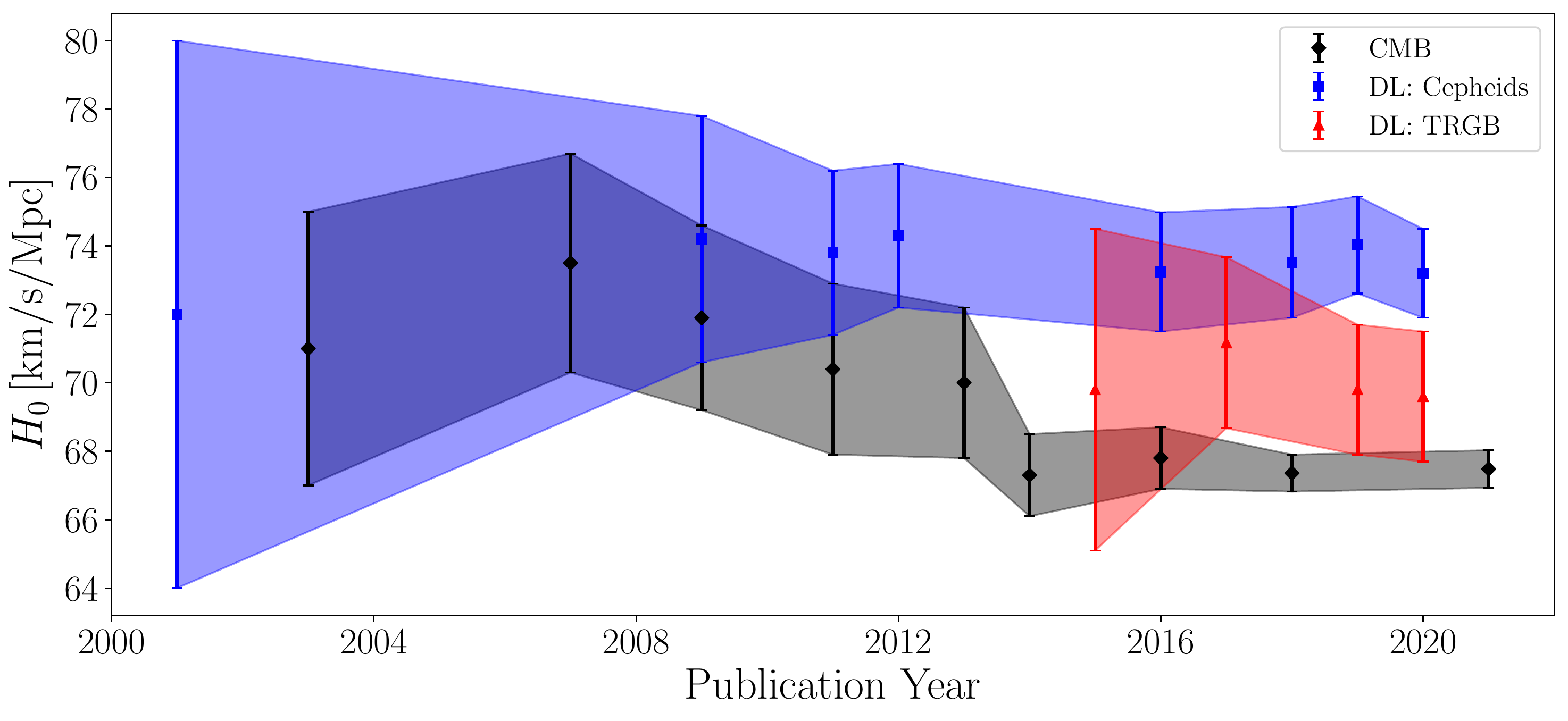}
\caption{A current status of the Hubble-Lema\^itre tension in early 2021. The black diamond error bars show the CMB measurements from WMAP, Planck, and SPT-3G, the blue square error bars show the distance ladder (DL) measurements based on cepheids, and the red triangle error bars show the DL measurements of $H_0$ based on the tip of the red giant branch (TRGB).}
\label{fig:tension}
\end{center}
\end{figure}

An illustration of the current status (as of early 2021) of the Hubble-Lema\^itre tension is shown in Figure \ref{fig:tension}, which is an update to a similar figure found in Freedman et al. (2019) \cite{Freedman_2019}. We clearly see there the first Planck CMB data release in 2013-2014 was instrumental in establishing the tension, with its statistical significance increasing with time as the error bars on the CMB and cepheid-based distance ladder measurements shrunk. More recently, distance measurements based on the tip of the red giant branch (at which the onset of core helium burning occurs) have lead to $H_0$ values that are somewhat intermediate between the CMB- and cepheid-based measurements, potentially calling into question the statistical significance of the tension. 

\section{Hubble Constant Measurements}\label{sec:hubble}
Fundamentally, every inference of the Hubble constant is always based on first establishing an \emph{absolute} distance scale, from which other distances can be compared. For CMB- and BAO-based inferences of the Hubble constant, this scale is the physical size of the baryon-photon sound horizon at the baryon drag epoch \cite{2019ApJ...874....4A}. For the distance-ladder-based inference of $H_0$, this absolute scale could be obtained via geometric parallax distances to Milky Way cepheids or red giant branch stars, or the distance to a nearby galaxy measured using detached eclipsing binaries \cite{2019Natur.567..200P} or via maser emission \cite{Pesce:2020xfe}. Needless to say that the value of the Hubble constant inferred from any technique is only as accurate as the value of this absolute calibrator scale.

\subsection{Cosmic microwave background and baryon acoustic oscillation}\label{subsec:cmb_bao}
The hot hydrogen and helium plasma in the pre-recombination era of our Universe had a relativistic sound speed, hence allowing density waves to propagate over cosmological distances at early times. These waves  propagated until free electrons and protons combined to form neutral atoms, allowing the photons to finally free stream out of baryonic density fluctuations and forming what we now know as the CMB. With their sound speed plummeting due to the loss of photon pressure support, the density waves then become frozen into the distribution of baryonic gas. As the gas cools and start forming stars and galaxies, the memory of these early plasma waves remains as they become imprinted into the large-scale distribution of galaxies. This phenomenon is what we now refer to as the baryon acoustic oscillation (BAO) since it causes small oscillations to appear in the galaxy power spectrum.

From this sound-wave picture, we expect the temperature of CMB photons to be strongly correlated on angular scales corresponding to how far the plasma sound wave could have travelled before the epoch of recombination. This angular scale is given by
\begin{equation}\label{eq:theta_s}
\theta_* = \frac{r_{\rm s}}{D_{\rm A}(z_*)},\quad {\rm where} \quad r_{\rm s} = \int_{z_*}^\infty \frac{c_{\rm s} dz}{H(z)}\quad{\rm and}\quad D_{\rm A}(z_*) = \int_0^{z_*} \frac{dz}{H(z)}.
\end{equation}
Here, $r_{\rm s}$ is the baryon-photon sound horizon, $z_*$ is the redshift at which baryons and photons decouple, $c_{\rm s}$ is the baryon-photon sound speed, $H(z)$ is the Hubble expansion rate as a function of redshift $z$, and $D_{\rm A}(z_*)$ is the angular diameter distance to $z_*$. Since $D_{\rm A}(z_*)$ is sensitive to the expansion rate at late times, we have $D_{\rm A}(z_*)\propto H_0^{-1}$. On the other hand, the sound horizon is sensitive only to the Hubble expansion rate in the pre-recombination epoch and is thus independent of $H_0$. We thus schematically have $\theta_*\propto H_0 r_{\rm s}$. While detailed measurements of the CMB anisotropies yield a highly precise value of $\theta_*$ \cite{Aghanim:2018eyx}, we see that measuring $H_0$ from the CMB requires us to know $r_{\rm s}$. Similarly, BAO observations are sensitive to the ratio of the comoving sound horizon $r_{\rm s}$ to the average angular diameter distance to the sample of galaxies used for the measurements, $\theta_{\rm BAO}  = r_{\rm s}/D_{\rm A}(z_{\rm gal}) \propto H_0 r_{\rm s}$. Being a three dimensional measurement, we can also observe the BAO signal for galaxies located at a similar sky location but different redshifts, but this measurement is likewise sensitive to the product $ H_0 r_{\rm s}$. This brings us to our first important lesson of this mini review: {\bf The inference of the Hubble constant from CMB and BAO data is only as good as our knowledge of the baryon-photon sound horizon.}

From Eq.~(\ref{eq:theta_s}) above, we see that $r_{\rm s}$ depends on the baryon-photon sound speed, the Hubble expansion rate in the pre-recombination era, and the epoch $z_*$ at which decoupling happens. The value of $z_*$ is largely determined by standard atomic physics and the COBE FIRAS \cite{1999ApJ...512..511M} measurement of the current CMB temperature, while $c_{\rm s}$ depends primarily on the baryon-to-photon ratio.  On the other hand, $H(z)$ depends on the energy content of the Universe at early times, which in the standard $\Lambda$-Cold-Dark-Matter ($\Lambda$CDM) model is made of photons, neutrinos, dark matter, and baryons. Within this model, the rich structure of the CMB temperature and polarization anisotropy spectra provides constraints on the abundance of all these components, hence allowing us to determine the actual value of our calibrator scale and thus inferred $H_0$. Of course, different assumptions about the energy budget of the early Universe could affect our calculation of the sound horizon, which we discuss more below. 

\subsection{Distance ladder}\label{sec:DL}
Compared to the aforementioned case of the baryon-photon sound horizon which requires several assumptions to be determined, the absolute calibrator scale for the distance-ladder inference of the Hubble constant is based on simple trigonometric geometry. The simplest example of this is the parallax distance to nearby stars, which can be simply determined by looking at the change in sky position of said stars as compared to distant background stars and quasars as the Earth orbits around the Sun. With this absolute distance scale in hand, a measurement of the photon flux from this star reaching our telescopes here on Earth can be converted (via the $1/r^2$ flux law) to its absolute luminosity. Knowing this, if we were able to observe the \emph{exact same star} placed at a different distance from us, we could immediately determine that distance by simply measuring the flux hitting our telescopes. In the real Universe, while two stars are never exactly the same, there exist stellar objects which have nearly constant absolute luminosities no matter where they are in the sky. These so-called \emph{standard candles} are key to establishing the distance ladder required to infer $H_0$. 

Of particular importance, both historically and for the current Hubble tension, are cepheid variable stars whose luminosity is known to vary in a periodic way. The pioneering work of Henrietta Leavitt \cite{1912HarCi.173....1L} showed that there is a tight relationship between their oscillation periods and their maximum detected flux. Once properly calibrated using the aforementioned parallax distances to nearby cepheids, this Leavitt law can be used to turn period measurements of distant cepheids into absolute luminosity measurements, which in turns tell us the absolute distance to these variable stars. This technique can be used to measure the distance to nearby galaxies but does not on its own tell us anything about the Hubble constant. Indeed, measuring the expansion rate today of course requires us to probe objects that \emph{are in the Hubble flow.} The problem is that galaxies that are in the Hubble flow are quite distant from us, making it difficult to directly observe cepheid variables in them. In other words, cepheids are just too faint to be seen at cosmological distances. 

Fortunately, the Universe has provided us with another type of standard candles that are extremely bright and can be easily seen at cosmological distances from us: type Ia supernovae. While we know that these are standard candles due the specific physical conditions leading to these extraordinary explosions, we still need to calibrate them to determine their absolute luminosity. This is where the cepheids and TRGB stars become handy: if we could measure a type Ia supernova in a nearby galaxy that also host cepheids or TRGB stars, we could use the latter to infer the absolute distance to this galaxy, and then use this distance to compute (again using the $1/r^2$ flux law) the absolute luminosity of the supernova. This is the essence of the cosmological distance ladder. Now, nearby galaxies hosting both cepheids and type Ia supernovae are rare: 19 of these calibrator galaxies were known as of Riess et al.~(2019) \cite{Riess:2019cxk}. This number is expected to double in the near future. In addition, another set of calibrator galaxies can be obtained from the TRGB method, further strengthening this rung of the distance ladder.

Equipped with the absolute luminosity (or magnitude in astronomy jargon) of type Ia supernova, we can use  the measured flux from distant supernovae in the Hubble flow to immediately determine their luminosity distances $d$ away from us. Then, measuring their receding velocities $v$ via their redshifts, we can determine the Hubble constant via Hubble's law, $H_0 = v/d$ (modern analyses use more precise modeling of the expansion, but this doesn't affect our main message here).This means that the inference of the Hubble constant does not come from $z=0$ observations. This brings us to our second main lesson of this mini review: {\bf the distance-ladder measurements of the $z=0$ expansion rate ($H_0$) is based on information from higher redshifts (typically $z\sim 0.02-0.15$).} This has immediate consequence on the type of solutions that can successfully address the tension, which we discuss briefly below. Fundamentally, the distance ladder does not measure $H_0$ directly; what it actually measures is the absolute luminosity of type Ia supernovae, which is ultimately tied to the absolute geometric measurement (from parallax or others) used to anchor the ladder. Ignore this at your own peril. 

\section{Interpretation and Solutions}\label{sec:solutions}

The above discussion points to two possible interpretations of the Hubble-Lema\^itre tension. On the one hand, one could blame the apparent discrepancy between distance ladder and CMB-BAO inferences of the Hubble constant on an incorrect value of the baryon-photon sound horizon. On the other hand, one could portray the tension as an issue with the inferred value of the absolute luminosity of type Ia supernovae.

\subsection{A sound horizon problem}
Since CMB and BAO data are primarily sensitive to the parameter combination $ H_0 r_{\rm s}$, increasing the value of the Hubble constant inferred from these data to match that determined from the cepheid-calibrated distance ladder requires us to \emph{decrease} the size of the sound horizon. Perhaps the most straightforward way to suppress the size of the sound horizon is to increase the Hubble expansion rate in the pre-recombination epoch, which requires us to increase the energy density at that time. This suppresses the integrand of the sound horizon in Eq.~(\ref{eq:theta_s}), hence yielding a smaller value. The challenge here is to shrink the sound horizon without ruining the complete CMB temperature and polarization spectra, which have by now be measured with exquisite precision. In particular, the CMB is sensitive to the behavior of the gravitational potentials at early times, which themselves depend on the different energy components populating the Universe. Adding new exotic energy components or increasing the energy density of known components will invariably affect the behavior of the potentials, and thus modify the CMB anisotropy spectra. The CMB is also sensitive to length scales other than the sound horizon (in particular, the photon diffusion scale), which is general have different dependency on the Hubble rate at these early epochs. Finally, changing the expansion rate at early times affects not only the CMB but also the evolution of matter fluctuations, which can be probed via weak gravitational lensing among other techniques. Despite these challenges, successful models have been proposed and work continues to determine whether they can be made compatible with all present and future cosmological data.

\subsection{An issue with type Ia supernova luminosity}
Another way to interpret the Hubble-Lema\^itre tension is that the type Ia supernova luminosity inferred from the distance ladder is too dim. A higher luminosity would place the supernovae further away from us for a given redshift, resulting in a lower inferred Hubble constant. A plausible culprit is of course the underlying geometric absolute distance measurements underpinning the whole distance ladder. As it turns out, there appears to be significant discrepancies \cite{Efstathiou:2020wxn,Efstathiou:2021ocp} between different geometric distance measurements, including (but not exclusively) between distances inferred from cepheids and TRGB stars.  More work (and data!) is needed to explore whether this could be a viable interpretation of the current tension, Anyone interested in the Hubble-Lema\^itre tension should pay close attention to work on this front in the next few years.

\section{The Fallacy of Late Solutions}
We conclude by mentioning that popular ``late solutions'' that aim at rapidly increasing the Hubble expansion rate close to $z=0$ do not actually address the root cause of the tension \cite{Benevento_2020,Efstathiou:2021ocp}. This is because calibrated type Ia supernovae fundamentally tell us about their luminosity distances from us, which depends on the integrated expansion history and not just on $H_0$. In other words, simply increasing $H(z)$ as $z\to0$ does not necessarily lead to a redshift-distance relation that is compatible with that inferred from the distance ladder.

\section*{Acknowledgments}
This work was supported in part by the US National Science Foundation through grant AST2008696. I  thank Barry Madore and George Efstathiou for useful comments. 

\section*{References}
\bibliography{ref}

\end{document}